# Memory Effects in Scattering from Accelerating Bodies

V. Kozlov[1,a)], S. Kosulnikov[1], D. Vovchuk[1,2] and P. Ginzburg[1]

[1]S*chool of Electrical Engineering, Tel Aviv University, Tel Aviv, 69978, Israel*

[2]*Department of Radio Engineering and Information Security, Yuriy Fedkovych Chernivtsi National University, Chernivtsi, 58000, Ukraine*

**Abstract:**

Interaction of electromagnetic, acoustic and even gravitational waves with accelerating bodies forms a class of nonstationary time-variant processes. Scattered waves contain intrinsic signatures of motion, which manifest in a broad range of phenomena, including Sagnac interference, Doppler and micro-Doppler frequency shifts. While general relativity is often required to account for motion, instantaneous rest frame approaches are frequently used to describe interactions with slowly accelerating objects. Here we investigate theoretically and experimentally an interaction regime, which is neither relativistic nor adiabatic. The test model considers an accelerating scatterer with a long-lasting relaxation memory. The slow decay rates violate the instantaneous reaction assumption of quasi-stationarity, introducing non-Markovian contributions to the scattering process. Memory signatures in scattering from a rotating dipole are studied theoretically, showing symmetry breaking of micro-Doppler combs. A quasi-stationary numeric analysis of scattering in the short memory limit is proposed and validated experimentally with an example of electromagnetic pulses interacting with a rotating wire.

---

a) vitaliko@mail.tau.ac.il

**Introduction**

Wave scattering theory is at the forefront of numerous disciplines, including classical and quantum mechanics [1], [2], electromagnetism [3], and even gravitation [4]–[6]. The advances made in these fields, inspired numerous applications such as medical imaging, radar and sonar, to name just few. Scattering from static bodies is one of the most explored scenarios, owing to the relative mathematical simplicity of the analysis as well as the ability to describe a broad range of real-life observations [7]. Yet the world remains insistently time dependent, where the location of objects, their form and even material composition are subject to time variations. This time dependence is responsible for an emergence of a broad range of well-known and understood effects, most notably Sagnac interference [8] and Doppler frequency shift, which is produced by uniformly moving scatterers [9]. However, scattering scenarios become much more complicated in the case of accelerated bodies, requiring the use of relativity [10]–[12]. As a result, both the form of the equations and the associated boundary conditions change, making scattering problems extremely difficult for analysis and interpretation. Nevertheless, significant simplifications can be made for nonrelativistic motion, where adiabatic (quasi-stationary) approaches can achieve accurate results. In this case, the instantaneous rest frame assumption considers a series of static configurations, resembling a discrete-time path of an object. This approach allows solving a large number of static problems, reproducing time-dependent effects by stitching the results sequentially. *The question to be asked in this context, is whether there is an overlooked regime of scattering, which does not require using the complex tools of relativity on the one hand, but cannot be considered with straightforward adiabatic approaches?*

Here we investigate a new and previously unexplored regime of wave phenomena, introducing memory effects into the scattering processes. Moving electromagnetic resonators with long relaxation times are proposed as a test model. Specifically, the analytically solvable problem of scattering from a rotating dipole is investigated. While the following investigation may be somewhat biased towards electromagnetic scattering, the concepts of a dipole and memory are universal, making the results relevant to many wave related disciplines.

The manuscript is organized as follows: Theoretical derivations for the induced moment in rotating dipoles with and without memory are performed and compared. Then a quasi-stationary memoryless method of performing time dependent numeric simulations with arbitrarily modulated incident waves is proposed, allowing complex time dependent problems to be solved using off-the-shelf static solvers. Finally, this method is utilized for solving an exemplary problem in electromagnetic scattering, where incident pulsed radiation is scattered from a rotating wire. Incident pulses of various lengths compared to the rotation frequency of the dipole are considered, revealing the importance of another kind of memory, the initial angle of the wire, which affects the scattered spectrum.

**General Discussion**

Since wave equations are often linear (or can be approximated as such), the scattering processes can be conveniently viewed through the prism of linear system analysis [13]. In this picture, static scattering could be envisioned as a linear time invariant (LTI) process, where the input into the system is an incident field and the output is, for example, the scattered field. It is well known that LTI systems cannot generate new spectral frequencies, therefore such effects as the Doppler shift cannot be described in the frame of this analysis. Instead, linear time variant (LTV) systems should be considered [14]. An illustration of the concept is depicted on Fig.1. Indeed, motion is the source of the Doppler shift, meaning that the scattering scene must be time dependent and described by LTV systems. In contrast with an LTI system, the output of an LTV system is no longer the time convolution between the input and the system's impulse response. The latter is often called Green's function in the context of a scattering. This renders Fourier techniques unsuitable for straightforward analysis of LTV systems, forcing a time domain approach. In the following, we shall consider a very specific LTV, which can provide intuitive insight into the mathematical treatment and consequences of memory effects that would follow.

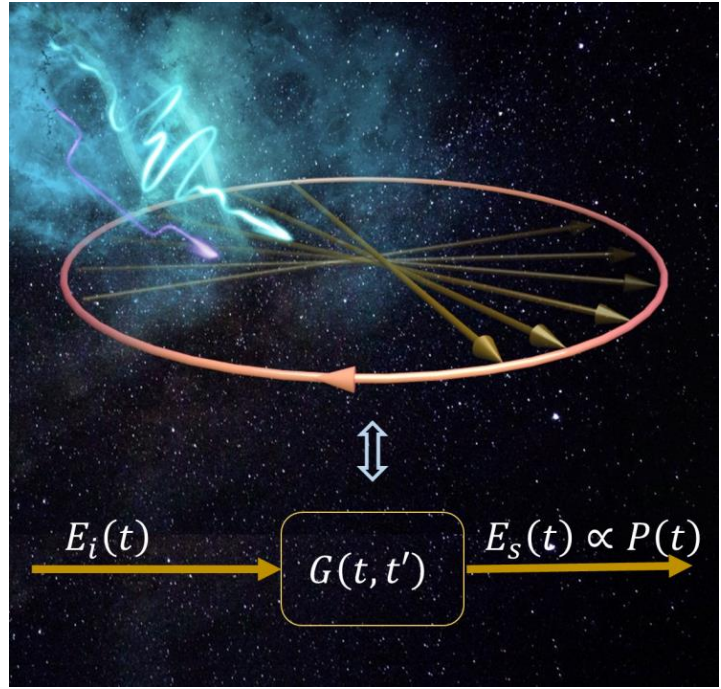

Figure 1: Illustration of memory effects on wave-matter interaction in accelerating reference frames. A specific example is an incident pulse scattered from a rotating dipole with memory. Slow relaxation rates permit the excitation of the dipole at one moment to decay when the scatterer had changed its position significantly. The scattering process may be thought of as a linear time variant process, where non-Markovian behaviour manifests in nontrivial signatures at the scattered far field.

As a test problem, let us consider a small subwavelength dipole (in the electromagnetic case, a short metallic wire), rotating around its center (Fig 1). The input to this LTV system is an incident transverse wave, polarized in the plane of the dipole's rotation, while the output of the system is defined as the induced dipole moment. For a stationary dipole, pointing along the polarization direction of the incident wave, the polarization mismatch is minimal and therefore the induced dipole is maximal. A perpendicularly aligned dipole, on the other hand, has maximal polarization mismatch and therefore there would be no induced dipole at all. Now let us allow the dipole to rotate with a constant angular frequency and find, what would be the induced dipole moment at any given time?

It is convenient to consider an ultra-short pulse which is exciting the rotating dipole. The excitation will inevitably decay as the dipole returns the induced energy back into the field. If this decay (radiation reaction) is fast compared with the time scale of the motion, it is tempting to apply time scale separation, which means assuming that there are two different times associated with the scattering process. The first is the so-called fast time, which is on the order of the carrier frequency period, while the second is the slow time of the scatterer's motion. This separation allows treating the slow time as a parameter, essentially assuming that the scatterer is momentarily static. The scattering problem is repeatedly solved in the fast time (for example by using Fourier techniques), allowing the slow time to advance forward. The resulting slow-time envelope of the dipole moment is therefore the 'stitching' of static solutions together. In particular, when the dipole is perpendicular to the polarization of the incident field, its moment will be identically zero, as is the case for the static dipole. This time scale separation method, also referred to as quasi-stationarity or adiabatic [10], [11], is reminiscent of similar approaches, such as the slowly varying amplitude approximation and short time Fourier transform (STFT) methods [15]. These approximations are especially prevalent in applied fields such as radar and sonar [16]–[19], where the typical velocity of scatterers is far smaller than that of the velocity of the impinging waves. This time-scale separation approach, when applicable, is a powerful tool for numeric simulations, as it allows the use of off-the-shelf static simulators in order to solve time dependent problems (see for example [20]–[22]). However, as all approximations do, they also must have inaccuracies and scenarios where they fail. The primary goal of this investigation is to quantify the deviation of quasi-stationary approximations when scattering memory is introduced and reveal the underlying physical processes that are being neglected when memory effects are not considered.

In order to understand the failure of time scale separation, consider the case where the excitation of the rotating dipole decays *slower* than the speed of the characteristic motion of the scatterer, i.e. the system possesses a long memory. In particular, the dipole could be excited by an ultra-short pulse exactly when it is aligned with the polarization direction, causing maximal moment excitation. This excitation might not decay by the time the dipole had rotated to the perpendicular position and therefore a non-zero induced moment would be observed at the time related to the perpendicular position, in stark contrast with the quasi-stationary result obtained earlier. This suggests that the ratio between the memory depth of the system, which is the reciprocal of the decay rate, and the characteristic time-scale of motion is crucial to the scattering process. This memory effect can still be significant even if the speed of the scatterer is far below the relativistic limit, provided that the system has a long enough memory depth. This will be shown mathematically ahead.

**Induced moments of rotating dipoles, exact versus quasi-stationary solutions**

Consider a generic problem of scattering from a subwavelength thin wire, treated by using a dipolar approximation. In the rest frame, the differential equation governing the dipole moment $P(t)$ is LTI, since all the coefficients that multiply the temporal derivatives are time independent (Eq. 1). In this case, the evolution of the moment is solely governed by initial boundary conditions, resonant frequency $\omega_0$, a driving force $F(t)$ which is related to the incident wave $E_i(t)$ and decay rate $\gamma$, which contains both radiation as well as dissipative losses. It is noted that the resonant frequency and decay rates are determined by the geometry and material make-up of the dipole in question (see for example [23]–[25] for a rigorous treatment). In the case of a short wavelength and lossless wire dipole, it holds true that $\omega_0 \propto \frac{1}{l}$ where $l$ the length of the wire and $\gamma \propto R_{rad}$, which is the radiation resistance. The differential equation for the dipole moment is therefore:

$$\ddot{P}(t) + 2\gamma\dot{P}(t) + \omega_0^2 P(t) = F(t). \qquad (1)$$

While the equation is quite general, a range of parameters and their relation to physical observables will be discussed after the experimental section. Eq. 1 can be solved either in the frequency or time domains, yet for the purpose of this investigation, it is instructive to choose the time domain, which serves to develop intuition about the memory effect. The solution of Eq. 1 is given by:

$$P(t) = \int_{-\infty}^{t} G(t - t')F(t')dt', \qquad (2)$$

where $G(t - t')$ is termed here as Green's memory function, and it is given by

$$G(t-t') = \begin{cases} \frac{1}{\sqrt{\omega_0^2 - \gamma^2}} e^{-\gamma(t-t')} Sin\left(\sqrt{(\omega_0^2 - \gamma^2)}(t-t')\right), t \geq t' \\ 0, otherwise \end{cases} \quad (3)$$

The intuitive interpretation of Eq.2 is that the dipole moment at some time $t$ depends on the excitation at earlier times $t'$, weighted by the decaying Green's memory function. As seen in Eq. 3, the depth of the memory, i.e. how far in the past can previous excitations affect the present, is determined by the decay rate $\gamma$. For large $\gamma$ the system possesses very little memory while for small $\gamma$ the system retains the memory of past excitations for a longer duration of time.

Now, consider the same dipole, rotating at a constant angular frequency $\dot{\theta}$ while being subjected to a resonant and linearly polarized incident plane wave with harmonic temporal dependence of $E_i(t) = Cos(\omega_0 t)$, where $\omega_0$ is the carrier frequency. The polarization mismatch factor $Cos(\theta(t))$ depends on the angle at any moment between the dipole orientation and the incident field polarization, turning Eq.2 into the following causal relation:

$$P(t) = \int_{-\infty}^{t} \underbrace{G(t-t')Cos(\theta(t'))}_{\tilde{G}(t,t')} E_i(t')dt', \quad (4)$$

where $\theta(t) = \theta_0 + \dot{\theta}t$ and $\theta_0$ is the initial angle between the dipole's principal axis and the polarization of the incident field at time 0. From the perspective of linear systems, when defining the input as the incident field and the output as the induced dipole moment, Eq.4 suggests that the system is LTV, which can be verified by observing that the associated Green's memory function $\tilde{G}(t, t')$ is no longer a function of the time difference $t - t'$. This is intuitively explained following similar logic to the one offered in the introduction, where the induced dipole depends strongly on the exact time the incident wave impinged upon the wire. By choosing $\theta_0 = 0$ for simplicity (this angle will only be important in the final section), the integral in Eq.4 may be solved exactly:

$$P(t) = \frac{1}{2\gamma^2(1+\Omega^2)} \sum_{\pm} \frac{\left(\Omega^2 \pm 2\Omega\sqrt{Q^2-1} - 1\right)Cos(\omega_\pm t) - \left(2\sqrt{Q^2-1} \pm \Omega\right)Sin(\omega_\pm t)}{3 - 4Q^2 \mp 4\Omega\sqrt{Q^2-1} - \Omega^2}, \quad (5)$$

where $Q \equiv \frac{\omega_0}{\gamma}$ is the Q-factor of the resonator, $\Omega \equiv \frac{\dot{\theta}}{\gamma}$ is the rescaled dimensionless rotation frequency and $\omega_\pm = \omega_0\sqrt{1 - Q^{-2}} \pm \dot{\theta}$ are the frequencies of the excited dipole moment. The summation in Eq.5 is to be understood as taking first the top value of $\pm$ (or $\mp$), and then adding the same expression with the bottom value. It is now possible to take the single sided Fourier transform of Eq.5 to reveal the intensity and phase of the induced dipole spectrum. Clearly, there are two peaks at frequencies $\omega_\pm$, with complex amplitudes

$$\tilde{P}_\pm = \sqrt{\frac{\pi}{8}} \frac{1}{\gamma^2(1+\Omega^2)} \frac{\left(\Omega^2 \pm 2\Omega\sqrt{Q^2-1}-1\right) - 2i\left(\sqrt{Q^2-1} \pm \Omega\right)}{3-4Q^2 \mp 4\Omega\sqrt{Q^2-1}-\Omega^2}. \qquad (6)$$

Before discussing the implications of Eq.6, it is instructive to consider the quasi stationary solution, obtained under the time-scale separation approximation. By observing Eq.4 it is tempting to assume that the frequency of the polarization mismatch $Cos(\theta(t))$ varies in time on a much slower scale in comparison with the incident field and Green's memory function $G(t-t')$. Moreover, if Green's memory function is fast decaying compared to the rate of the polarization mismatch (i.e. $\Omega = \frac{\dot\theta}{\gamma} \ll 1$), it effectively turns into a sampling function at time $t' = t$. This memoryless property permits the removal of the polarization mismatch term outside of the integral, leading to:

$$P_{memoryless}(t) = Cos(\theta(t)) \int_{-\infty}^{t} G(t-t')E_i(t')dt'. \qquad (7)$$

The form of Eq.7 consists of an integral, which is the solution to the static problem at minimal polarization mismatch, multiplied by a time-dependent amplitude related to the instantaneous position of the dipole. This result indicates the required assumptions, which allow investigating short memory systems by means of static solutions with an additional slow-time dependent multiplicative parameter. This equation is also the basis for the last part of the manuscript, where a method for solving dynamic problems with static numeric solvers will be presented and discussed. The solution of Eq.7 is immediate:

$$P_{memoryless}(t) = -\frac{1}{2\gamma^2} \frac{Cos(\omega_\pm t) + 2\sqrt{Q^2-1}Sin(\omega_\pm t)}{3-4Q^2}. \qquad (8)$$

The single sided Fourier transform of Eq.8 provides complex amplitudes of the induced dipole moment spectrum, which reveals the existence of two peaks, similar to what was obtained in the exact solution of Eq.6:

$$\tilde{P}_{memoryless} = -\sqrt{\frac{\pi}{8}} \frac{1}{2\gamma^2} \frac{1+2i\sqrt{Q^2-1}}{3-4Q^2}. \qquad (9)$$

It is easy to conclude that Eq.5 and Eq.6 reduce to Eq.8 and Eq.9 respectively when the rotation is slow enough (i.e. $\Omega = \frac{\dot\theta}{\gamma} \ll 1$). The major difference between the memoryless system and the exact one can be seen by comparing the amplitudes of the spectral components - while the peak magnitudes are equal in the memoryless case (Eq.9), the long memory scenario is characterized by asymmetric amplitudes (Eq.6). This behavior is shown in more detail on Fig.2(a). The magnitude and phase of the peaks in Eq.6 and Eq.9 can be plotted versus the dimensionless rotation frequency $\Omega$, as shown in Fig.2(b) and Fig.2(c). The exact form of the curves depends on the quality factor Q of the resonator. For high Q, there is very little asymmetry at slow rotation rates. Yet, when the dimensionless rotation frequency increases to $\Omega = 2Q$, there is maximal asymmetry as the $\tilde{P}_+$ spectral peak continues it's fast decay to zero, while $\tilde{P}_-$ returns to the value of the memoryless system. While this result is quite interesting, it is important to note that the condition for this strong asymmetry is $\dot{\theta} = 2\omega_0$, which might be an extremely fast rotation rate for optical or microwave incident wave frequencies. For such fast rotations, relativity must be taken into account along with other radiative corrections and the classical framework of the obtained results is no longer valid. This resonant regime also means that the mechanical motion of frequency $\dot{\theta} = 2\omega_0$ pumps the dipolar excitation frequency of $\omega_0$, in similarity with the relativistic case explored in [26]. It is possible to observe this asymmetric spectral amplitude effect for waves that have low frequencies, as may be the case for mechanical, as well as for very low frequency (VLF) electromagnetic waves, which are used for underwater communication [27], [28]. In addition, these memory effect could be observed for reactively loaded dipoles that were made with the purpose of obtaining resonant properties with extremely small apertures. Such dipoles are constricted by the Chu-Wheeler limit [29], [30] and can possess extremely small decay rates. This is further explored in the conclusions, where a compact formula for the required reactive load at a given dipole length is derived, showing that this effect can be observed even for modest rotation periods if large reactance's are used.

For low Q the amplitude of the peaks in Eq. 9 behaves very differently due to the detuning of the resonant frequency of the rotating dipole from $\omega_0$ to $\omega_0\sqrt{1 - Q^{-2}}$. While the positive spectral peak at frequency $\omega_+$ still decays from the memoryless solution as it were in the high Q case, the negative spectral amplitude $\tilde{P}_-$ actually exceeds the value of the memoryless rotating dipole, even at low rotation speed. While the amplitude asymmetry is more prominent in the low Q case, for the phase of the spectral peaks the result is opposite. This effect occurs since the high Q case shows faster deviation rate between the induced phases of the spectral peaks at low rotation frequencies.

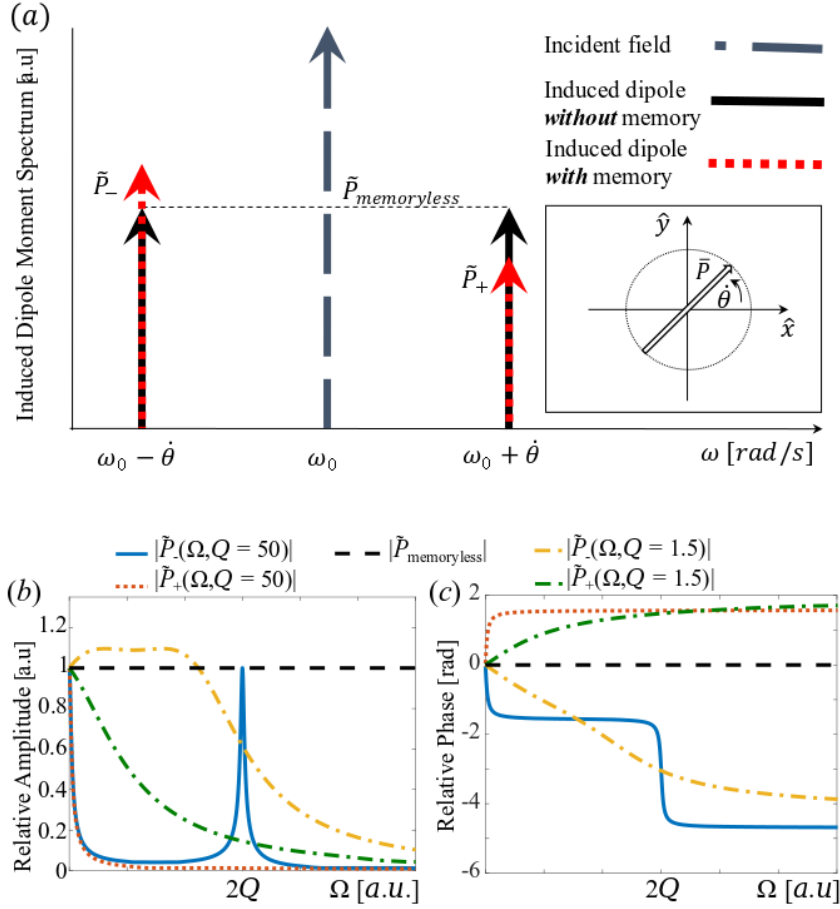

Figure 2: Comparison between the exact and quasi-stationary (adiabatic, time scale separation) solutions. (a) The spectrum of the induced dipole moment contains two frequencies. The memoryless solution (Eq.6) has equal amplitudes and phases while the scattering from a dipole with memory (Eq.9) has an asymmetric spectrum. Inset – a dipole, rotating with angular frequency $\dot{\theta}$. (b) and (c) show the behavior of the amplitude and phase of the spectral peaks, for high and low Q-factors, as a function of dimensionless rotation frequency $\Omega = \frac{\dot{\theta}}{\gamma}$.

It is straightforward to calculate the scattered field once the dipole moment is known (see for example [31]–[33]) and will not be explicitly developed here. It will be sufficient to say that in the rest frame of the dipole, the point of scattered field observation is also moving periodically with the rotational frequency, meaning that the scattered field will have frequency components of $2\omega_{\pm}$ in the lab frame [34]. The mathematical derivation presented in this section is in fact more general and applies to more than rotating dipoles alone. When moving from Eq.2 to Eq.4, the rotation of the dipole introduced polarization mismatch, only to be later removed from the integral as per Eq.7. But the polarization mismatch is not unique and in fact it could have been the modulation of the incident field carrier that underwent the same analytic process, this will be expanded on ahead (see Eq.10). Furthermore, since the incident modulation could be arbitrarily time dependent, obtaining the solution for the harmonic case (constantly rotating dipole with polarization mismatch, as described above) allows using Fourier series in order to obtain the solution to any type of modulation. The correctness of this solution will depend on the highest significant harmonic contained in the Fourier series of the carrier modulation, where the condition for the correctness of the quasi-stationary solution will be that this maximal frequency is required to be slower than the memory depth of the system. This insight will be used in the next section to synthesize a method for using static numeric simulators to compute dynamic time dependent scattering problems in memoryless, or very short memory systems.

**Quasi-stationary method for numeric time dependent simulation of short memory systems**

Analysis of scattering from rotating bodies has quite a few applied aspects. Micro-Doppler signatures, being the manifestation of internal degrees of mechanical motion within an object, are the subject of studies in radar and sonar sciences. Micro-Doppler spectra, which are unique characteristics of the observed scatterer undergoing complex motion, are used for remote targets' identification. For example, rotating blades of a helicopter or a drone [35], [36], jet engine modulation [37], cyclists and pedestrians [38] are among numerous examples, where micro-Doppler spectroscopy brings an advantage in identification and classification via careful signal post-processing. The vast majority of signals probe micro-Doppler with either ultra-short or quasi-CW (continuous wave) signals. In the first case of short pulses, the target can be assumed static during the interaction, meaning that the carrier frequency of each wave packet remains unaffected. The Doppler effect in that case manifests in the phase difference between consecutive pulses. In the second case of quasi-CW, the pulse length is effectively longer than the characteristic time scale of the motion, hence the phase and amplitude modulation is imposed on the carrier frequency directly. In this case it is more instructive to look at the micro-Doppler frequency comb in the frequency domain [20], [21], [34], which is the spectral content within the quasi-CW pulse. In both cases, time-scale separation techniques are frequently applied for the scattering analysis [16]. Here we analyse and experimentally demonstrate the transition between a micro-Doppler spectrum to a discrete frequency comb by tuning the duration of the probing pulse. When the pulse length is smaller than the period of dipole rotation, a different kind of memory is revealed as the initial state of the system – which in this case is the angle between the incident polarization and the dipole (wire) orientation at the time of pulse incidence. The LTV property in this case comes through boundary conditions instead of the response kernel.

Straightforward simulation of moving scatterers, is not trivial, since the vast majority of electromagnetic numeric software cannot handle scattering problems with a moving media self-consistently. Static simulations, on the other hand, are very common and can be performed with numerous off-the-shelf software packages available (CST, Comsol, Lumerical and HFSS to name a few). At this point it is important to note that in the considered practical scenario the decay rate of the excited current on the dipole is extremely fast, with the whole scattering process taking less than a nanosecond. Since the rotation speed of the wire in the experiment that will be presented ahead does not exceed 10Hz (i.e. $\Omega = \frac{\dot{\theta}}{\gamma} \ll 1$), it is clearly possible to apply the time scale separation method, suggested in Eq.7. The way to perform a time dependent simulation is therefore by creating a sweep on the position of the wire, solving a static scattering problem at every location, and then stitching these solutions together with a slow time parameter (sequential counter of positions). Simulating the modulation of the incident carrier is nontrivial as well, as it significantly increases the computational effort. There are two ways to go about simulating the scattering process of short pulses, where the carrier's modulation plays a role. The brute force method is to evaluate the scattering at a number of specific frequency points, which correspond to the spectral shape of the modulation, and then adding them up coherently with Fourier weights that govern the magnitude and phase of each frequency within the desired modulated spectrum. Finally, the weighted sum can be put through an inverse Fourier transform in order to obtain the time domain scattered field (carrier and modulation). The drawback of such a method is that it requires the evaluation of numerous frequency points, which can be computationally heavy, especially when many scatterer positions need to be calculated. In order to perform the analysis in a more computationally efficient way, a different approach is suggested here for memoryless systems, which is the case for the majority of practical scattering scenarios. As per the discussion at the end of the last section, Eq.7 deals with the removal of the polarization mismatch from the integral. The polarization mismatch is not unique, and the same applies to slow modulation of the carrier, under similar restrictions. The result is a static solution multiplied by a time dependent amplitude. Consider Eq.4, but this time with an incident Gaussian pulse of the form $E_i(t) = Cos(\omega_0 t)A(t)$, where $A(t)$ is a slowly varying amplitude, and is of a Gaussian shape in the simulations and experiment ($A(t) = e^{-\left(\frac{t}{\tau}\right)^2}$, $\tau$ is the characteristic width of the pulse). By using Eq. 4 together with the quasi-stationary approximation of Eq.7, which is correct for slow rotation and modulation (i.e. $\frac{\dot{\theta}}{\gamma}, \gamma\tau \ll 1$), the dipole moment is given by:

$$P(t) \approx Cos(\theta_0 + \dot{\theta}t)A(t)\underbrace{\int_{-\infty}^{t} G(t-t')Cos(\omega_0 t')dt'}_{static\ harmonic\ solution}, \qquad (10)$$

This result could be intuitively understood since the dipole moment would be proportional to the amplitude of the incident pulse at any given moment, multiplied by the static response. The method was applied to the problem of pulse scattering from a rotating dipole as shown in the experiment on Fig.3. The simulation was performed using CST Studio, where the wire was illuminated by a plane wave, while the structure was rotated by a set of sufficiently discretized angles, corresponding to the location of the wire at any time during the motion. The far field at the receiving antenna location (perpendicular to the direction of incident wave propagation) was recorded for each angle. The resulting slow time sequence was then multiplied by the modulation envelope, as shown on Fig.4. Note, that an additional slow time-dependent rotation operator should multiply Eq. 10 to account for the transformation of scattered fields from the rotating lab frame to the laboratory one. It is interesting to note that for pulses shorter than the rotation frequency of the dipole (i.e. $\dot{\theta}\tau \ll 1$), the resulting dipole moment, and therefore the scattered field, will be proportional to $Cos(\theta_0)$. This is because in that case the scattered pulse effectively samples the momentary position of the wire. This memory effect of initial position will be shown experimentally ahead to be of considerable importance even when the pulse length is on the order of rotational period (i.e. $\dot{\theta}\tau \sim 1$). The results of the simulation appear ahead on Fig.4, Fig.5 and Fig.6, where they are compared against the experiment, which is discussed next.

**Experimental verification of quasi-stationary method - scattering from a slowly rotating wire**

In order to verify the validity of the numerical technique, a set of experiments were performed in an anechoic chamber. The setup is shown on Fig.3, where a rotating wire of length $L_{wire} = 40$ mm and diameter $d_{wire} = 1$ mm serves as the analogue of a rotating dipole. The wire is rotated by a stepper motor at a constant frequency of 3.4Hz while being illuminated by vertically polarized electromagnetic field. The transmitted signal is a Gaussian envelope around a 3.195 GHz carrier, which is chosen to maximize the scattering from the wire. The Gaussian modulation has a controllable temporal width to satisfy different interaction regimes, which will be described below. The receiving antenna is polarized in the vertical direction as well and is placed perpendicularly to the transmitter, in order to increase the isolation between transmitting and receiving channels, while simultaneously insuring direct lines of sight to the scatterer. This bi-static configuration allows increasing signal to noise ratio quite significantly. The signal from the receiving antenna is downconverted by using a pair of mixers and a 90° splitter (maintaining the phase in one path and adding 90° in another one), producing in phase and quadrature data at the intermediate frequency (IF) output of the mixers. The IF outputs are recorded using a sampling scope and the data is digitally processed in order to obtain the down converted temporal signal.

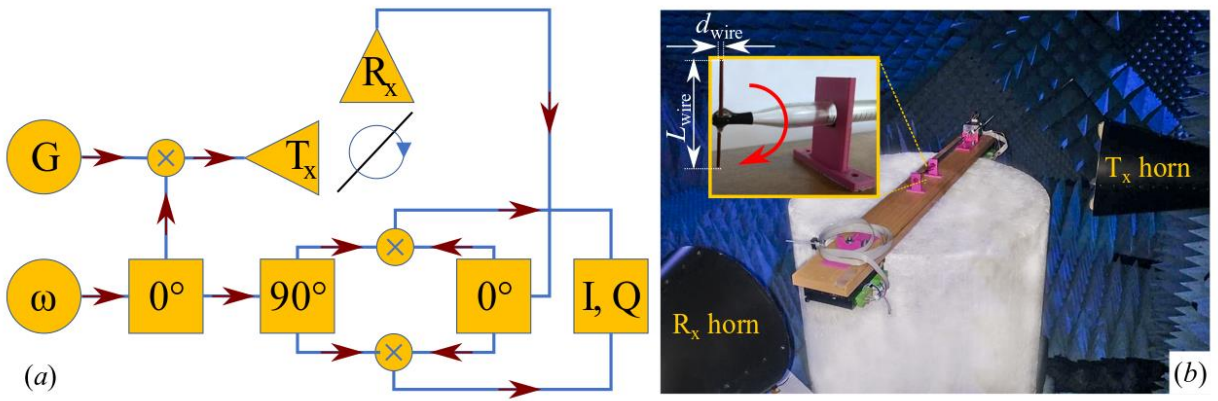

12the modulation envelope, as shown on Fig.4. Note, that an additional slow time-dependent rotation operator should multiply Eq. 10 to account for the transformation of scattered fields from the rotating lab frame to the laboratory one. It is interesting to note that for pulses shorter than the rotation frequency of the dipole (i.e. $\dot{\theta}\tau \ll 1$), the resulting dipole moment, and therefore the scattered field, will be proportional to $Cos(\theta_0)$. This is because in that case the scattered pulse effectively samples the momentary position of the wire. This memory effect of initial position will be shown experimentally ahead to be of considerable importance even when the pulse length is on the order of rotational period (i.e. $\dot{\theta}\tau \sim 1$). The results of the simulation appear ahead on Fig.4, Fig.5 and Fig.6, where they are compared against the experiment, which is discussed next.

**Experimental verification of quasi-stationary method - scattering from a slowly rotating wire**

In order to verify the validity of the numerical technique, a set of experiments were performed in an anechoic chamber. The setup is shown on Fig.3, where a rotating wire of length $L_{wire} = 40$ mm and diameter $d_{wire} = 1$ mm serves as the analogue of a rotating dipole. The wire is rotated by a stepper motor at a constant frequency of 3.4Hz while being illuminated by vertically polarized electromagnetic field. The transmitted signal is a Gaussian envelope around a 3.195 GHz carrier, which is chosen to maximize the scattering from the wire. The Gaussian modulation has a controllable temporal width to satisfy different interaction regimes, which will be described below. The receiving antenna is polarized in the vertical direction as well and is placed perpendicularly to the transmitter, in order to increase the isolation between transmitting and receiving channels, while simultaneously insuring direct lines of sight to the scatterer. This bi-static configuration allows increasing signal to noise ratio quite significantly. The signal from the receiving antenna is downconverted by using a pair of mixers and a 90° splitter (maintaining the phase in one path and adding 90° in another one), producing in phase and quadrature data at the intermediate frequency (IF) output of the mixers. The IF outputs are recorded using a sampling scope and the data is digitally processed in order to obtain the down converted temporal signal.

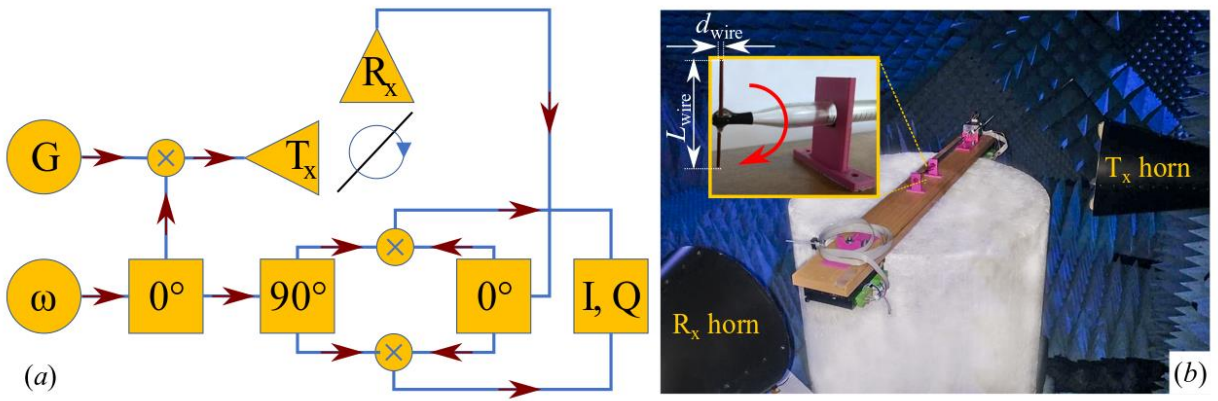

(a) (b)

Figure 3: Experimental set up for probing scattering from rotating objects. (a) the schematic representation of the setup. A carrier frequency $\omega$ is split between a transmitting and receiving arms. The transmitting arm is mixed with a Gaussian (G) pulsed signal of controllable pulse length to generate the envelope. The receiving arm is split into two quadratures in order to down-convert the scattered signal. (b) A photo of the experimental setup inside the anechoic chamber. Inset – the enlarged photograph of the mechanism to control the wire's rotation (electric dipole analogue, rotation direction is marked with red arrow) at a constant frequency of 3.4 Hz.

The results, comparing the theory and experiment, appear in Fig. 4, Fig.5 and Fig.6, and are in a good agreement with each other. There are several important factors to consider about the system under study. First, the relation between the pulse duration and the rotation period of the wire should be underlined. As discussed in the introduction, short pulses will sample the instantaneous location of the wire, while long pulses will experience periodic amplitude modulation. Second, in the short pulse regime the initial angle of the wire with respect to the polarization of the incident field plays a crucial role, as shown in Eq.10. Taking those clarifications into account, the simulations and experiments are repeated for pulse lengths ranging from 0.1s to 100s, as well as for initial angles of the wire $\theta_0$ ranging from 0 to 180°. Fig.4 shows the experimental and numerical results for two pulse widths and with the wire having the initial angle of $\theta_0 = 0$ (the position of the wire at time t=0, when the pulse has its maximum at the wire's centre). The results are the time dependent (Fig.4a and Fig.4c), as well as the spectral (Fig.4b and Fig.4d) content of the down converted scattered field at the receiver. Indeed, it can be observed that for incident Gaussian pulses, the scattered field is also Gaussian, with additional harmonic modulation, as predicted by the theoretical result in Eq.10. For longer pulses, shown in Fig.4b, the spectrum is comprised of a frequency comb, which becomes narrower with increasing pulse length. Indeed for infinite pulse length the result converges into a discrete comb, which is separated apart in frequency by $2\dot{\theta}$ (as in [34], which considers CW radiation). For short pulse lengths shown on Fig.4d, the spectrum is smeared, as the components at each discrete frequency spread and overlap with neighbouring peaks.

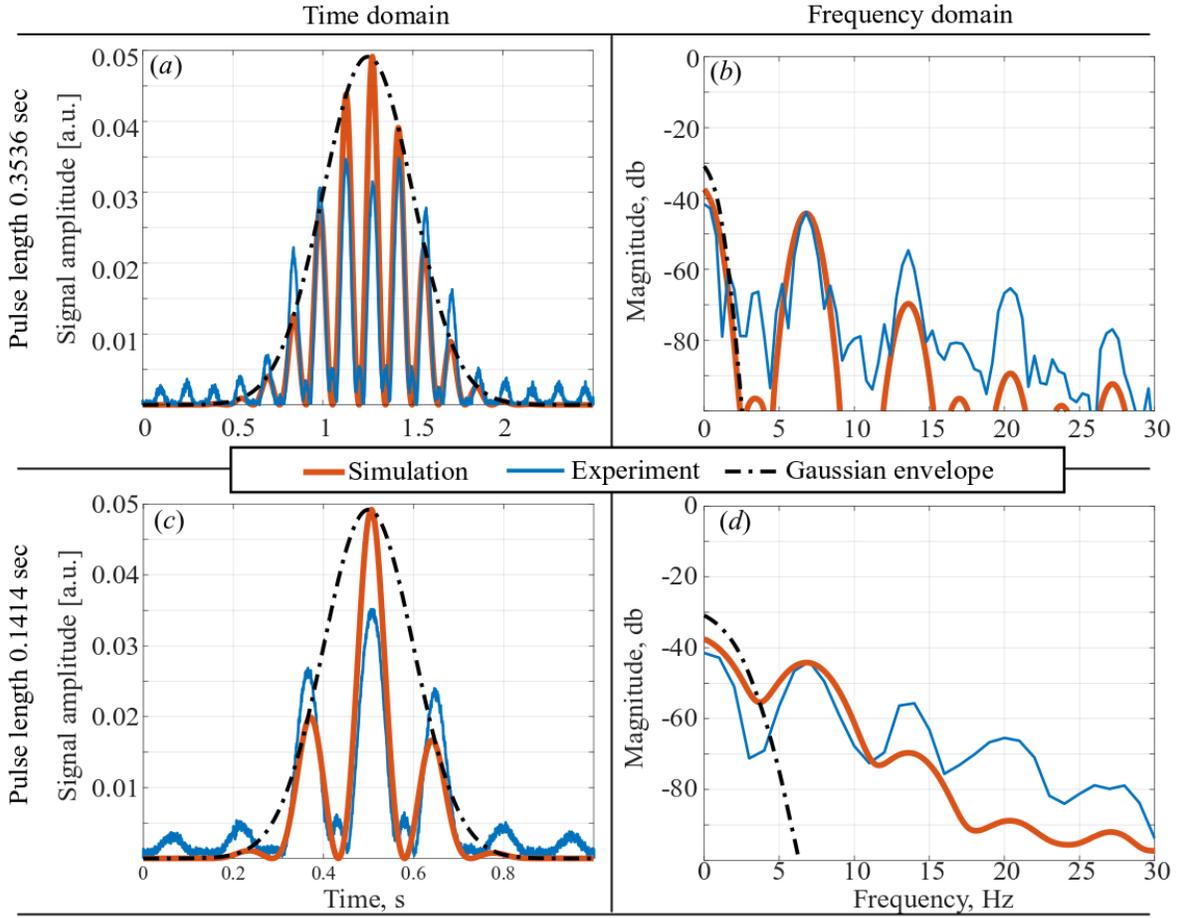

Figure 4: Far-field scattering of Gaussian pulses from a rotating wire, comparing short memory (adiabatic, quasi-stationary) simulation with experiment (the setup appears in Fig. 3). Blue lines - two different Gaussian pulse lengths, 0.3536Hz and 0.1414Hz measured experimentally. Red lines – numerical analysis using the short memory method. Black dashed-dotted line – the envelope of the incident pulse (left column) and its spectrum (right column). Columns – scattered signals in time and frequency domains respectively.

In order to perform a comprehensive study of the phenomena, a broad range of parameters have been investigated. In particular, the same experiment that produced Fig.4 was repeated with pulse lengths ranging from 0.1s to 100s, as well as for initial angles of the wire $\theta_0$ ranging from 0 to 180°. The results are shown on Fig.5 and Fig.6. Note that Fig4b and Fig.4c represent horizontal cuts of Fig.5, which reveal the transition from long to short pulse regime, where smearing of the Doppler comb is clearly seen. The discrete frequency lines (x-axis of the color map) become broader when reducing the pulse length (y-axis), until they begin overlapping and finally smearing out, once the rotational period of the dipole becomes comparable with the signal's Gaussian envelope. The experimental data contains additional peaks in
14

between the one's found in simulation. This is the result of offset of the rotation axis from the center of the wire, leading to odd harmonics generation at the baseband (discussed in more detail in [34]).

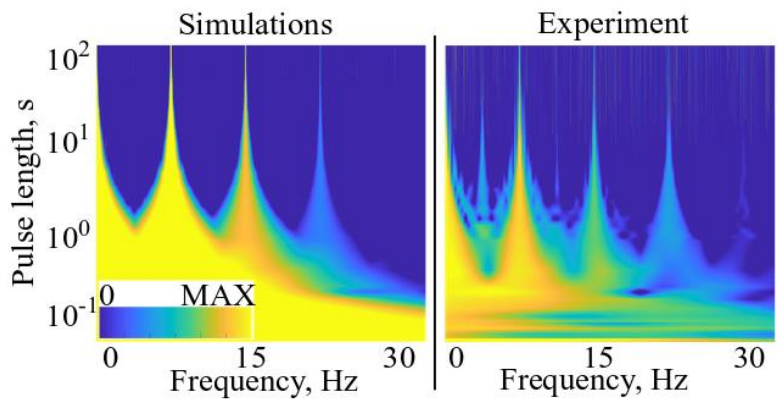

Figure 5: Comparison of the scattered baseband spectrums obtained in simulation and experiment as a function of frequency and pulse length. As the pulses length increases (in comparison with rotation speed), the micro Doppler peaks form at discrete frequencies. For short pulses, the peaks are broader in frequency, creating a featureless spectral continuum. The initial angle for both experiment and simulation is taken as $\theta_0 = 0$.

The final parameter affecting the interaction phenomenon is the initial angle between the dipole and the polarization of the incident radiation. This angle is defined in Eq.10 as the one between the incident polarization and the wire's principal axis at a time corresponding to the arrival of the peak of the incident Gaussian pulse to the origin. Fig.6 shows the spectral composition of the scattered pulses at fixed pulse lengths for variable initial angles. For pulses longer in comparison with the rotation period of the wire, there is no dependence of the spectrum on the initial angle, as shown on Fig.6a and Fig.6b. This is due to the fact that the wire completes numerous rotations during the duration of the pulse, effectively losing the memory of the initial angle. On the other side, when the pulse length is very short in comparison with the wire's rotation frequency, the spectrum is completely dominated by the initial angle. This can be understood by returning to Eq.10, where the short pulse effectively samples the cosine at time 0, causing a harmonic dependence of the scattered spectrum on the initial angle. As shown on Fig.6g and Fig.6h, for pulse length of 1.25s, there appears a minimum of the scattered spectrum near the initial angle of 90°. This is consisted with Eq.10 and can be understood intuitively, since the wire is oriented perpendicularly to the incident polarization and, hence, scatters very little. For extremely short pulses this minimum would occur at 90° exactly. However, for longer pulse lengths, the minima may have an offset, as shown in Fig.6g and Fig.6h, where the pulse length is only about twice smaller than the rotation period. The sensitivity of the spectral peaks to the memory of initial angle allows for remote sensing of the state of the rotating wire. In Fig.4b, Fig.4c, Fig.5 and Fig.6 only the positive frequencies of the baseband are shown, since the negative frequency components are identical due to the fact that this scattering scenario has very short memory (Eq.9) and hence, symmetric micro-Doppler spectrum. For faster wire rotation, or for another type of resonator (typically, with high Q-factor), the quasi-stationary method would fail, resulting in asymmetric micro-Doppler peaks (Eq.6). In this case, double-sided maps should be presented for a complete characterization of the process.

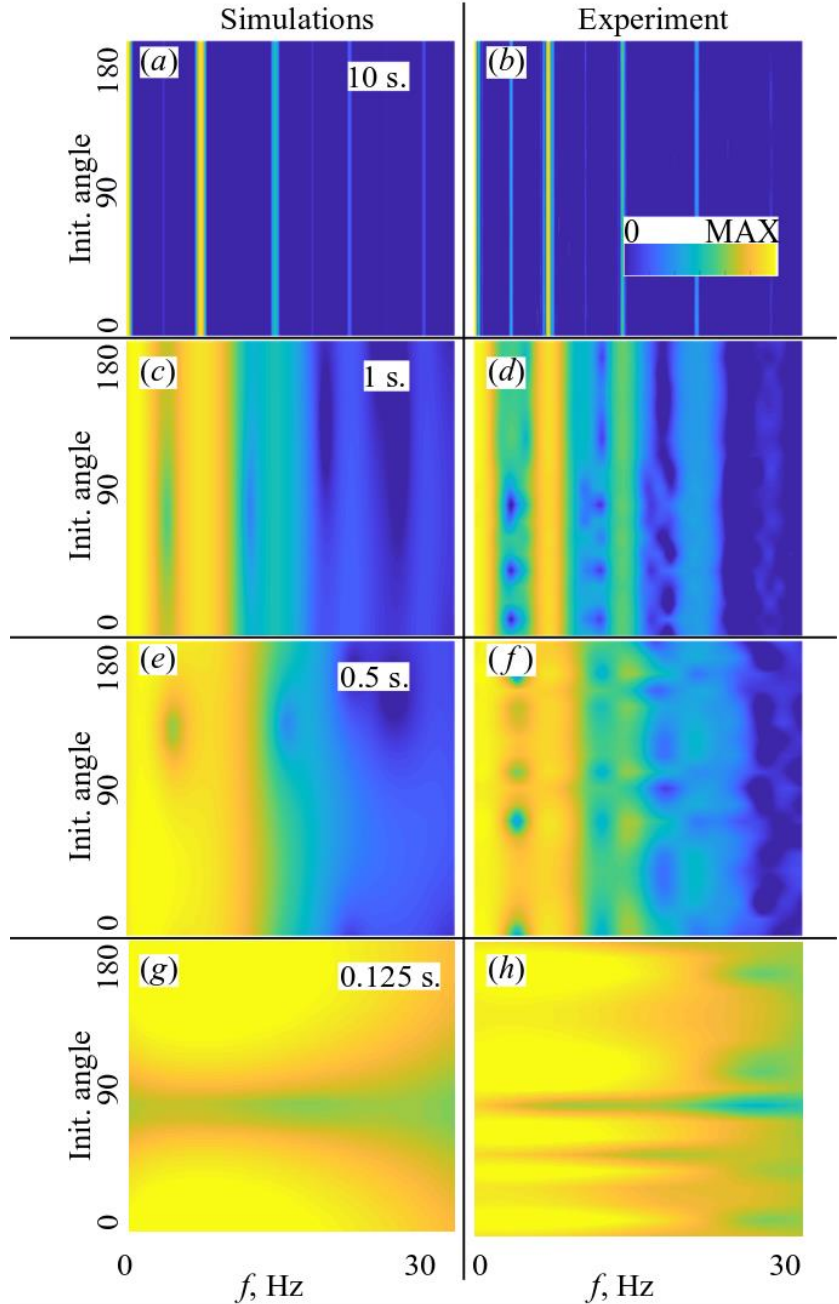

Figure 6: Colormap, demonstrating scattered signals spectra at the baseband. Vertical axis – initial angle between the wire and the polarization of the incident field, horizontal axis – frequency. Different pulse widths are indicated in insets. Numerical and experimental data are in left and right columns, correspondingly.

**Summary and Discussion**

A new wave-matter interaction regime of motion-induced non-stationarity was investigated. In particular, scattering from rotating dipoles was shown to have strong dependence on memory effects, which manifests in asymmetric micro-Doppler signatures of the scattered field. When a scatterer does not possess intrinsically long memory, an adiabatic (quasi stationary) simulation method was suggested and experimentally tested, allowing off-the shelf static numeric solvers to be used to compute time dependent scattering scenarios. While straightforward experimental demonstration of the memory-inspired impact on scattering is still technologically challenging, other aspects of nonstationarity in moving systems were probed. The impact of initial boundary conditions and pulse duration, in comparison with the characteristic parameter of motion, were investigated, demonstrating the transition between micro-Doppler combs and continuous spectra. It is also worth noting that the incident pulse shape affects the scattering scenario, as the Fourier spectral decomposition in nonstationary problems has limited physical meaning, seeing as the output of an LTV system is no longer the usual convolution between the input and impulse response of the system. For example, two signals, having the same spectrum but different in phase content, might have a completely different micro-Doppler spectrum.

An experimentally feasible configuration, where the proposed scenario can fully emerge, can be achieved by reactively loading the rotating dipole in the radio frequency range. The first resonant frequency of a dipole corresponds to the condition where its length $l$ is at about half an excitation wavelength, which is related to the angular frequency $\widetilde{\omega}_0 \simeq \frac{\pi c}{l}$. To shift this resonance towards lower frequencies while keeping the dipole length constant, the dipole may be reactively loaded, for example with a capacitor in parallel to the natural capacitance of the dipole. Recalling that the dipole may be modeled as an RLC contour with resonant frequency $\omega_0 = \frac{1}{\sqrt{LC}}$, where $L$ and $C$ are the combined natural and lumped element reactance, and by considering the newly added reactive capacitor as possessing capacitance which is $\beta$ times larger than the natural one, the resonance is down shifted to $\omega_0 = \frac{\widetilde{\omega}_0}{\sqrt{1+\beta}} = \frac{\pi c}{l\sqrt{1+\beta}}$. Since memory effects begin to manifest themselves when $\dot\theta \approx \gamma$, we supplement the beforehand expression with the Chu-Harrington limit, applied on small scatterers and get: $\frac{\omega_0}{\gamma} \geq \left(\frac{2c}{\omega_0 l}\right)^3 + \frac{2c}{\omega_0 l} \stackrel{\beta \gg 1}{\approx} \left(\frac{2c}{\omega_0 l}\right)^3$, therefore the interesting regime is reached when $\dot\theta \approx \frac{\pi^4 c}{8l\beta^2} = \frac{3.65\left[\frac{GHz}{meter\ of\ wire}\right]}{\beta^2}$. For a 1m dipole which is 5mm in diameter, loaded with a realistic capacitor of coefficient $\beta = 10^3$ (which for the above values means about a 1nF lumped capacitor [52]), the resonant frequency is shifted towards $\omega_0 = 4.74 MHz$. The required rotational frequency in order for the memory effects to begin playing a role in the scattering process is only $\dot{\theta} = 3.65 KHz$, being achievable with modern instrumentation.

The investigated memory effect is universal and can emerge in many wave-related disciplines and scenarios, e.g. optical, where molecules and even macroscopic structures are being rotated with laser beams [39],[40],[41],[42],[43] and optofluidics, where time-dependent Purcell enhancements can emerge [44],[45]. In radar, LIDAR and sonar, where the micro-Doppler signatures produced by rotating blades of a helicopter are of interest [34], [46], [47], and even in astronomy, where neutron stars can be approximated by rotating dipoles [48]. While the effects of rotation on scattering were comprehensively studied in the past (e.g. [34], [49]–[51]), the effects of memory and nonstationarity were not considered before, to the best of the authors knowledge. The new interaction regime was studied here primarily in the context of the Doppler effect, however memory effects can be important in other scattering phenomena as well.

**Acknowledgments**

The research was supported in part by ERC StG "In Motion" ((802279)) and PAZY Foundation (Grant No. 01021248).